\documentclass[useAMS,usenatbib]{mnras}
\usepackage{times,graphicx,amssymb,amsmath,natbib}

\usepackage{hyperref}
\hypersetup{draft}
\usepackage[pdftex]{}
\pdfoutput=1
\newcommand{\be}{\begin{equation}}
\newcommand{\e}{\end{equation}}
\newcommand{\bear}{\begin{eqnarray}}
\newcommand{\ear}{\end{eqnarray}}

\newcommand{\f}{\frac}
\newcommand{\de}{{\rm d}}

\defcitealias{2015ApJ...813L...8M}{MH15}
\defcitealias{2007ApJ...654..731H}{H07}
\defcitealias{2015A&A...578A..83G}{G15}

\def\nat{Nature}

\def\prd{Phys. Rev. D}
\def\mnras{MNRAS}
\def\apj{ApJ}
\def\apjl{ApJL}
\def\apjs{ApJS}
\def\aap{A\&A}

\def\araa{ARA\&A}
\def\aj{AJ}

\def\physrep{Phys. Rep.}
\def\apss{Ap\&SS}               
\def\pasp{Publ. Astr. Soc. Pac.}

\def\pasj{Pub. Astron. Soc. Japan}

\def\jcap{JCAP}

\begin{document}

\title[Reionization in closed $\Lambda$CDM inflation]
{First study of reionization in the Planck 2015 normalized closed $\Lambda$CDM inflation model}
\author[Mitra, Choudhury \& Ratra]
{Sourav Mitra$^1$\thanks{E-mail: hisourav@gmail.com},~
Tirthankar Roy Choudhury$^2$~
and
Bharat Ratra$^3$~\\
$^1$Surendranath College, Department of Physics, 24/2 M. G. Road, Kolkata 700009, India\\
$^2$National Centre for Radio Astrophysics, TIFR, Post Bag 3, Ganeshkhind, Pune 411007, India\\
$^3$Department of Physics, Kansas State University, 116 Cardwell Hall, Manhattan, KS 66506, USA
} 

\maketitle

\date{\today}

\begin{abstract}
We study reionization in two non-flat $\Lambda$CDM inflation 
models that best fit the Planck 2015 cosmic microwave background anisotropy
observations, ignoring or in conjunction with baryon acoustic oscillation 
distance measurements. We implement a principal component analysis (PCA) to 
estimate the uncertainties in the reionization history from a joint 
quasar-CMB dataset. A thorough Markov Chain Monte Carlo analysis is done
over the parameter space of PCA modes for both non-flat $\Lambda$CDM inflation 
models as well as the original Planck 2016 tilted, spatially-flat 
$\Lambda$CDM inflation model. Although both flat and non-flat models can 
closely match the low-redshift ($z\lesssim6$) observations, we notice a 
possible tension between high-redshift ($z\sim8$) Lyman-$\alpha$ emitter 
data and the non-flat models. This is solely due to the fact that the 
closed models have a relatively higher reionization optical depth 
compared to the flat one, which in turn demands more high-redshift ionizing 
sources and favors an extended reionization starting as early as $z\approx14$.
We conclude that as opposed to flat-cosmology, for the non-flat cosmology models
(i) the escape fraction needs steep redshift evolution and even unrealistically
high values at some redshifts and (ii) most of the physical parameters require
to have non-monotonic redshift evolution, especially apparent when Lyman-$\alpha$
emitter data is included in the analysis.
\end{abstract}


\section{Introduction}

Measurements of the cosmic microwave background (CMB) anisotropy by the 
Planck satellite tightly constrain cosmological parameters \citep{2016A&A...594A..13P}. 
Their results are consistent with the standard spatially-flat $\Lambda$CDM 
inflation model \citep{1984ApJ...284..439P} whose leading current-epoch 
constituents are dark energy ($\sim69\%$) in the form of a cosmological 
constant ($\Lambda$) and non-baryonic cold dark matter (CDM) ($\sim26\%$). 
Six parameters are needed to describe this standard model, namely the 
physical baryonic density parameter ($\Omega_{\rm b}h^2$, where $h$ is the 
Hubble constant $H_0$ in units of 100 km/s/Mpc), the physical CDM density 
parameter ($\Omega_{\rm c}h^2$), the angular size of the sound 
horizon at recombination ($\theta$), the reionization electron scattering 
optical depth ($\tau_{\rm el}$), and the slope ($n_s$) and amplitude ($A_s$) of 
the (assumed) power-law primordial scalar energy density inhomogeneity
power spectrum. Although the simple six-parameter tilted, spatially-flat 
$\Lambda$CDM
model has proven to be successful on most observational fronts \citep{2016A&A...594A..13P},
some challenging issues still remain unsettled. For example, 
the uncertainty in the nature of dark energy persists till date
\citep{1988ApJ...325L..17P,1988PhRvD..37.3406R,2000IJMPD...9..373S,2003PhR...380..235P,
2004LNP...653..141S,2007MNRAS.380.1573S,2008PASP..120..235R}. Another issue is that local
measurements of the expansion rate result in a higher $H_0$
\citep[e.g.,][]{2016ApJ...826...56R} than many other techniques
\citep{2011PASP..123.1127C,2012PhRvD..86d3520C,2013JCAP...10..060S,2015PhRvD..92l3516A,
2016A&A...594A..13P,2017JCAP...01..015L,2017ApJ...835...86C,2016A&A...595A.109L,
2017ApJ...849...84W,2017PhRvD..96h3532L,2017arXiv171100403D,2018ApJ...856....3Y}.

Recently, it has been argued that a closed $\Lambda$CDM (or XCDM or $\phi$CDM) 
inflation model could partially alleviate two possible drawbacks of the tilted 
spatially-flat $\Lambda$CDM model \citep{2017arXiv170703452O,2017arXiv171003271O,2017arXiv171208617O}.
In the closed $\Lambda$CDM inflation model that best fits the Planck 2015
CMB anisotropy data, the predicted CMB temperature anisotropy angular power spectrum,
$C_\ell$, where $\ell$ is multipole number, has less power at low $\ell$, in  better agreement
with the observations. Also, the resulting fractional energy 
density inhomogeneity averaged over 8$h^{-1}$ Mpc radius spheres, $\sigma_8$, is in 
better accord with lower estimates from weak lensing measurements. Both of 
these results are the consequence of the suppression of large-scale energy 
density inhomogeneity power in the best-fit closed inflation cases relative 
to the best-fit flat inflation model 
\citep{2017arXiv170703452O,2017arXiv171003271O}. However, the large $\ell$
$C_\ell$'s in the flat model provide a somewhat better fit to the observations 
than do those in the non-flat cases.

Nonzero spatial curvature provides an additional cosmological length scale so 
it is physically inconsistent to use a power-law energy density inhomogeneity 
power spectrum in a non-flat model. In a non-flat cosmological 
model inflation provides the only known way to compute the power spectrum. When 
the open inflation \citep{1982Natur.295..304G,1994ApJ...432L...5R,1995PhRvD..52.1837R}
and closed inflation \citep{1984NuPhB.239..257H,1985PhRvD..31.1931R,2017arXiv170703439R}
model energy density inhomogeneity power spectra are used to 
analyze the Planck CMB anisotropy data \citep{2016A&A...594A..13P}, they favor 
a closed Universe with current spatial curvature density parameter of magnitude
of a percent or two \citep{2017arXiv170703452O,2017arXiv171003271O,2017arXiv171208617O}.     
    
More precisely, \cite{2017arXiv170703452O} have analysed a six-parameter 
non-flat $\Lambda$CDM inflation model, parameterized by
$\Omega_{\rm b}h^2, \Omega_{\rm c}h^2, \theta, \tau_{\rm el}, \Omega_{\rm k}$ and 
$A_s$ (with previously considered free parameter $n_s$ now replaced by 
the current value of the spatial curvature density parameter $\Omega_{\rm k}$) 
by exploiting Planck 2015 CMB anisotropy \citep{2016A&A...594A..13P} and 
baryon acoustic oscillation (BAO) distance measurements
\citep{2011MNRAS.416.3017B,2014MNRAS.441...24A,2015MNRAS.449..835R}.
They found that the existing data favour a slightly closed non-flat model 
with $\Omega_{\rm k}=-0.018\pm0.008$ (1-$\sigma$ confidence limits; C.L.) when 
constrained against Planck CMB TT + lowP + lensing data alone, and with 
$\Omega_{\rm k}=-0.008\pm0.002$ when the BAO data are included along with the 
Planck CMB measurements. In both cases, the resulting present day Hubble 
parameter $H_0$ and matter density parameter $\Omega_{\rm m}$ are compatible 
with most other data on these parameters.\footnote{For $\Omega_{\rm m}$ 
see \citet{2003PASP..115.1143C}.} It might be significant that many analyses 
based on a variety of different non-CMB data (including BAO, Type Ia 
supernovae apparent magnitude, Hubble parameter, growth factor, and 
gravitational lensing data, as well as various combinations thereof)
also do not rule out the non-flat 
models \citep{2015Ap&SS.357...11F,2014PhRvD..90b3012S,2014ApJ...789L..15L,
2016PhRvD..93d3517C,2016ApJ...829...61C,2016ApJ...828...85Y,2017JCAP...01..015L,
2017ApJ...835...26F,2016ApJ...833..240L,2017ApJ...838..160W,2017JCAP...03..028R,2018ApJ...856....3Y}.

However, \citet{2017arXiv170703452O,2017arXiv171003271O,2017arXiv171208617O} find an interesting 
deviation from the original Planck results in another important aspect of 
observational cosmology, the value of the reionization optical depth 
$\tau_{\rm el}$, which has a direct influence on the epoch of reionization 
(EoR)\footnote{For reviews on reionization, we point the reader to \cite{LoebBarkana01,
BarkanaLoeb01,2006ARA&A..44..415F,tirth06a,tirth09,2013ASSL..396...45Z,
2014PTEP.2014fB112N,2014arXiv1409.4946F,2016ASSL..423...23L}.},
because the transition from a neutral intergalactic medium (IGM) to an ionized 
one drastically increases the free electron contents that can Thomson 
scatter the CMB photons. For the tilted spatially-flat $\Lambda$CDM inflation 
model, Planck estimates $\tau_{\rm el}$ to be $0.066\pm0.012$ from Planck 2015 \citep{2016A&A...594A..13P} or $0.055\pm 0.009$
from Planck 2016 \citep{2016A&A...596A.107P}. The Planck flat-$\Lambda$CDM 
constraint points to an instantaneous reionization
occurring at mean redshift $z_{\rm reion}\approx8-9$ \citep{2016A&A...596A.108P}, which is compatible with reionization
by the observed population of galaxies, namely PopII stars \citep{2015ApJ...802L..19R,mitra4}.
A lower optical depth might also explain the rapid decrease in the number density of Ly$\alpha$ emitters (LAEs) detected
at $z\sim7$ which would have been in marginal tension with models having a 
relatively higher $\tau_{\rm el}$
\citep{2015MNRAS.446..566M,2015MNRAS.452..261C}. On the other hand, in the 
closed $\Lambda$CDM model \cite{2017arXiv170703452O} reckon $\tau_{\rm el}$
to be quite high, which could have a severe impact on reionization at higher 
redshifts. Thus an in-depth investigation is needed on this aspect in order 
to address the significant differences between
the higher-$z$ predictions for reionization in Planck 2016 normalized 
tilted flat-$\Lambda$CDM and Planck 2015 normalized closed-$\Lambda$CDM models.

This paper presents a first study of reionization in the non-flat $\Lambda$CDM 
inflation scenario. We put our emphasis on a detailed comparison between the 
flat and non-flat cosmological models. In the next section we briefly discuss 
the main features of our semi-analytical reionization model and the datasets 
used here to constrain it. We present our findings in 
Section~\ref{sec:results}, and finally conclude in Section~\ref{sec:conclusions}.

\section{Reionization model and datasets}
\label{sec:cfmodel}

The reionization model used here is based on the semi-analytical approach of
\cite{tirth05} and \cite{tirth06}.

In this model, the ionization state of the IGM is well-described by a 
multi-phase medium, a mixture of both ionized and neutral regions. The density 
distribution of the IGM is assumed to have a lognormal form at low 
densities, changing to a power law at high densities \citep{tirth05}.
The model accounts for the inhomogeneities in the IGM using a description 
similar to that of \citet{2000ApJ...530....1M} in which reionization ends 
once all the low-density regions are ionized \citep{tirth09}.
For simplicity we assume that all photons are absorbed shortly after
being emitted (this is commonly known as the ``local source'' approximation), 
which is a reasonable approximation\footnote{However it's been argued that,
although the ionizing emissivity computed using
the local source approximation asymptotically approaches the exact value computed by solving the full
cosmological radiative transfer equation towards higher redshifts, it can be significantly
too low at $z\lesssim4$ \citep{2013MNRAS.436.1023B}. Since most our conclusions are derived
from data at $z \gtrsim 5.5$, we do not expect this approximation to affect them significantly.}
for $z\gtrsim3$ when the mean free path of photons is much smaller than the Hubble radius
\citep{1999ApJ...514..648M,tirth09,2003ApJ...584..110S}.

The ionizing ultra-violet (UV) photon budget is assumed to be produced by normal PopII
stars and quasars. Many lines of evidence suggest that star-forming galaxies 
dominate the UV radiation background at earlier epochs, while quasars 
dominate only at later times due to the rapid decline in their abundances
beyond $z\simeq6$
(\citealt{2007ApJ...654..731H,2015ApJ...813L..35K,mitra5,2017MNRAS.468.4691D,2018MNRAS.473..227H};
but also see \citealt{2015ApJ...813L...8M,2016MNRAS.457.4051K} for quasar-only
reionization models).
The model also incorporates the impact of radiative feedback (which increases 
the minimum star-forming halo mass in the ionized regions) on reionization 
by altering the minimum circular velocity of halos that
are able to cool. The production rate of ionizing photons is computed from
\be
\dot{n}_{\gamma} = N_{\rm ion} n_b \f{\de f_{\rm coll}}{\de t},
\e
where $f_{\rm coll}$ is the fraction of matter that has collapsed into halos, 
obtained by using an appropriate halo
mass function, $n_b$ is the total  baryonic  number  density and $N_{\rm ion}$ is the number of
ionizing photons in the IGM per baryon in stars, which can be written as a product of the star formation
efficiency $\epsilon_*$, escape fraction $f_{\rm esc}$ of the ionizing photons 
escaping into the IGM and the specific
number of photons emitted per baryon in stars, $N_{\rm ion}=\epsilon_*f_{\rm esc}N_\gamma$ \citep{mitra3,mitra4}.
In reality, this parameter depends on halo mass and redshift. 
Unfortunately, we do not have a physically motivated model for this, due to 
our limited understanding of complex star formation processes.

Here we ignore any explicit dependence of $N_{\rm ion}$ on halo mass. However,
with the help of a principal component analysis (PCA), it is possible to 
include a redshift dependence \citep{mitra1,mitra2,mitra4}. The PCA technique 
has proven to be very useful in re-expressing a large number of (possibly) 
correlated variables in a new basis of a smaller number of uncorrelated 
variables without significant loss of information.\footnote{PCA has been 
widely used in various astrophysical and cosmological data analyses,
see, e.g., \cite{1999MNRAS.304...75E,2002MNRAS.332..193E,2003PhRvD..68b3001H,
2003PhRvL..90c1301H,2006MNRAS.372..646L,2008ApJ...672..737M,2010PhRvL.104u1301C,
2011A&A...527A..49I,2012MNRAS.421.3570G,2015PhRvD..91f3514M}.}

We start by assuming that $N_{\rm ion} (z)$ is an arbitrary function of $z$ and 
design the Fisher information matrix with help of a suitable fiducial model 
using the observed datasets of
 (i) the hydrogen photoionization rates $\Gamma_{\rm PI}$ in the range $2.4\leqslant z\leqslant6$ from
 \cite{2011MNRAS.412.1926W} and \cite{2013MNRAS.436.1023B}\footnote{The datasets have
 a mild dependence on the adopted cosmological parameters which has been taken account of in
 our work here.};
 (ii) redshift evolution of Lyman limit systems (LLS), $\de N_{\rm LL}/\de z$ over a wide
 redshift range ($0.36 < z < 6$) from the combined data points of \cite{2010ApJ...721.1448S}
 and \cite{2010ApJ...718..392P}; and
 (iii) reionization optical depth $\tau_{\rm el}$ using three different constraints -
 a) recent Planck 2016 data ($0.055\pm0.009$; flat $\Lambda$CDM model) from \cite{2016A&A...596A.107P},
 b) non-flat $\Lambda$CDM with Planck 2015 CMB (TT + lowP + lensing) data ($0.101\pm0.021$)
 and c) non-flat $\Lambda$CDM with Planck 2015 CMB + BAO data ($0.120\pm0.012$) from
 \cite{2017arXiv170703452O}\footnote{Although we use Planck 2016
 $\tau_{\rm el}$ data for the flat model and Planck 2015 CMB data for the non-flat models,
 we have checked and found that if we use 2015 $\tau_{\rm el}$ data for the flat case, which
 has slightly higher  value of $0.066\pm0.012$, the main conclusions remain the same.
 Also, in the non-flat case, more recent analyses based on using significantly
 more non-CMB data, than the few BAO data points \cite{2017arXiv170703452O} used,
 results in a smaller $\tau_{\rm el} = 0.112 \pm 0.012$ \citep{2018arXiv180100213P,2018arXiv180305522P}.}.
 The Fisher matrix thus contains information 
regarding the sensitivity of all the individual datasets on $N_{\rm ion}(z)$.
The fiducial model $N_{\rm ion}^{\rm fid} (z)$ should be chosen in such a way that it
can match all the observables at $z<6$ and also produce a $\tau_{\rm el}$ in the
acceptable range. For the flat model we have taken a constant $N_{\rm ion}^{\rm fid}=10$
which is suited to match the Planck data as seen in \cite{mitra4}. Unfortunately, this
simplest constant model does not work for the non-flat cases, since we require
larger contribution from early epoch sources in order to achieve higher $\tau_{\rm el}$
(also seen in \citealt{mitra1,mitra2}). $N_{\rm ion}^{\rm fid}$ should be higher at early epochs where PopIII
stars are likely to dominate and should smoothly transit to a lower value
($N_{\rm ion}^{\rm fid}=10$) at $z\lesssim6$ determined by the usual PopII stars
to produce a good match with all the observations considered in this work.
Although the derived parameters somewhat depend on the fiducial model chosen,
and the actual form of underlying {\it true} $N_{\rm ion}$ might be slightly different
from it, the final conclusions of this paper (presented later) would hold for
any $N_{\rm ion}^{\rm fid}$ which can produce at least a reasonable match with the
observables mentioned above.
We further set a prior on the neutral hydrogen fraction using robust 
constraints obtained from the Ly$\alpha$ forest observations of distant quasars
by \cite{2015MNRAS.447..499M} at $z\sim5-6$;  $x_{\rm HI}<0.11$ at $z=5.9$ and $x_{\rm HI}<0.09$ at $z=5.6$.
We kept all other cosmological parameters, 
corresponding to the different models, at their best-fit values as
obtained by \cite{2016A&A...594A..13P} for the flat model and by 
\cite{2017arXiv170703452O} for the non-flat cases. For clarity, we quote 
those in Table~\ref{tab:params}. This means that the uncertainties on our 
reionization predictions here are tighter than they really should be; to 
account for the uncertainties on the other five cosmological parameters 
will require a more involved analysis.

\begin{table}
\centering
\begin{tabular}{@{\extracolsep{\fill} } l c c c}
Parameter & Flat model & \multicolumn{2}{c}{Non-flat models}\\
& & TT+lowP+lensing & TT+lowP+lensing+BAO\\
\hline
\hline
$\Omega_{\rm m}$ & $0.3089$ & $0.32$ & $0.28$\\
$\Omega_{\rm b}h^2$ & $0.0223$ & $0.02304$ & $0.02302$\\
$\Omega_{\rm k}$ & --- & $-0.018$ & $-0.008$\\
$h$ & $0.6774$ & $0.6433$ & $0.6823$\\
$\sigma_8$ & $0.8159$ & $0.797$ & $0.819$\\
$n_s$ & $0.9667$ & --- & ---\\
\hline
\hline
\end{tabular}
\caption{List of the best-fit cosmological parameters for flat (from \citealt{2016A&A...594A..13P})
and non-flat $\Lambda$CDM models (from \citealt{2017arXiv170703452O}). We 
ignore uncertainties in these parameters in our analyses here.}
\label{tab:params}
\end{table}
 
 Once we have the Fisher matrix, we can deconstruct it into pairs of eigenvalues and eigenvectors
 (also known as principal components, PCs). The primary objective of PCA is the dimensionality reduction of our
 fiducial parameter space. This can be done by identifying the more accurately determined modes
 with smaller uncertainties, which in turn correspond to the eigenmodes associated with larger
 eigenvalues. This results in a relatively fewer number of PCs needed for the reconstruction of the 
 true $N_{\rm ion}(z)$. The other modes with larger uncertainties (or equivalently smaller eigenvalues)
 can be discarded at this stage without significant loss of information.
 We assume that PopII stars are the sole contributor of $N_{\rm ion}$; another 
stellar population, such as PopIII stars, might be expected to manifest itself
as evolution of $N_{\rm ion}(z)$ with redshift \citep{mitra1,mitra2}.

\section{Results: MCMC-PCA constraints}
\label{sec:results}

\begin{figure*}
\centering
  \includegraphics[height=0.5\textwidth, angle=0]{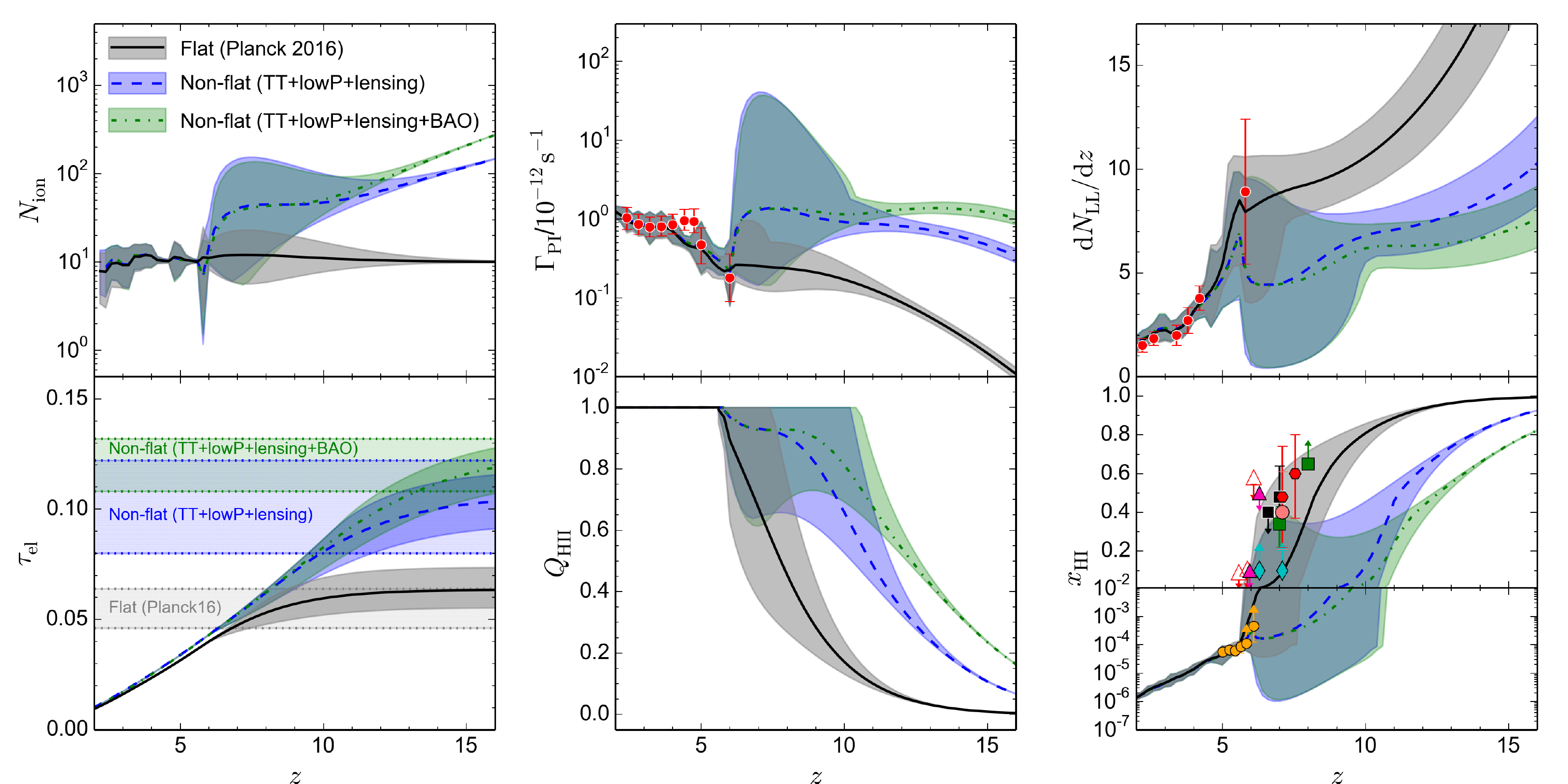}
  \caption{MCMC constraints on various  quantities  related
  to reionization history obtained from the PCA for three different cases: flat $\Lambda$CDM model with 
  Planck 2016 data; non-flat $\Lambda$CDM with Planck 2015 CMB (TT + lowP + lensing) data and non-flat $\Lambda$CDM with
  TT + lowP + lensing + BAO data. The lines correspond to the best-fit models 
  while the shaded regions correspond to their 2-$\sigma$ uncertainty ranges. 
  The red points with error bars denote the corresponding
  observational data points.
  {\it Top-left:} the evolution of the effective $N_{\rm ion}(z)$;
  {\it Top-middle:} the hydrogen photoionization rate $\Gamma_{\rm PI}(z)$ 
  along with observed data from \citet{2011MNRAS.412.1926W} and \citet{2013MNRAS.436.1023B};
  {\it Top-right:} the LLS distribution $\de N_{\rm LL}/\de z$ with combined data
  points from \citet{2010ApJ...721.1448S} and \citet{2010ApJ...718..392P}; 
  {\it Bottom-left:} electron scattering optical depth $\tau_{\rm el}$  and 
  constraints from \citet{2016A&A...596A.107P} and \citet{2017arXiv170703452O} (indicated by 
differently shaded regions for the three different cases);  
  {\it Bottom-middle:} the volume filling factor of HII regions $Q_{\rm HII}(z)$;
  {\it Bottom-right:}  the global neutral hydrogen fraction $x_{\rm HI}(z)$ with various current
  observational limits. We direct the reader to Figure~\ref{fig:xHI} for their references.}
\label{fig:MCMC}
\end{figure*}

Constraints on $N_{\rm ion}(z)$ and other quantities are obtained from Markov 
Chain Monte Carlo (MCMC) analyses over the relevant principal modes using the 
datasets mentioned above. We find that the first $2-7$ eigenmodes with largest 
eigenvalues suffice for this purpose. The uncertainties derived from each 
mode are combined to determine the total uncertainty in the final stage of 
reconstruction by using a model-independent Akaike information criterion \citep{2007MNRAS.377L..74L}.
For details see \cite{mitra1,mitra2,mitra4}. We repeat 
the whole analysis for all three cases considered here: flat $\Lambda$CDM 
and the two non-flat models with and without the BAO constraints.

The MCMC results are shown in Figure~\ref{fig:MCMC}. The colored shaded 
regions and the lines, with different styles for different cases, 
correspond to the 2-$\sigma$ (95\% C.L.) uncertainty ranges and mean values 
of those parameters, respectively,
obtained from MCMC statistics. All quantities are tightly constrained at 
$z\lesssim6$ as expected, due to the fact that most of the observed data 
related to reionization exist only at these redshifts. A wide range of 
histories at $z >6$ is still allowed by the data. The evolution at $z >6$
is essentially governed by the optical depth data alone, that's why a relatively
weaker constraint is apparent in this regime.  The 2-$\sigma$ C.L. also shows a decreasing trend at high redshifts
since the components of the Fisher matrix are zero as there exist no free electrons to contribute to
$\tau_{\rm el}$, providing no significant information from the PCs beyond this point.
The mean evolution of all the quantities for non-flat models is almost identical to the flat one at
$z\lesssim6$; at earlier epochs they differ significantly, as expected from the different electron scattering
optical depths. The overall 2-$\sigma$ errors at $z>6$ on all quantities for non-flat models are slightly
higher than those for the flat Planck 2016 model, as 
the observational uncertainty on the Planck 2016 $\tau_{\rm el}$ data is the 
lowest among the three models.

We find that, contrary to the flat Planck 2016 case, an evolving $N_{\rm ion}$ 
with redshift ({\it top-left panel})
is required for the non-flat Planck 2015 models due to higher values of $\tau_{\rm el}$. It is not possible to match
this $\tau_{\rm el}$ data with a constant $N_{\rm ion}$, i.e. $N_{\rm ion}$ must increase at $z >6$
for these models. This is a clear signature of either a changing initial
mass function (IMF) induced by chemical feedback from PopIII stars
and/or evolution in the star-forming efficiency and/or evolution in the 
photon escape fraction of galaxies.
These non-flat models show a relatively higher value of $\Gamma_{\rm PI}$ ({\it top-middle panel})
at early epochs than the flat model, as the former ones allow the contribution 
of ionizing photons from high-redshift PopIII stars. In fact, the PopIII 
photon contribution seem to be 
highest for the non-flat CMB + BAO case, as $\tau_{\rm el}$ for this model is 
the largest of all. A similar trend is also found in the evolution
of LLSs ({\it top-right panel}). In both panels we indicate the corresponding current observational constraints
(red points with error bars) at $z\lesssim6$ which we have included in this MCMC analysis.
All three models match these quite accurately. Another key quantity of interest is the volume filling factor
$Q_{\rm HII}(z)$ for ionized hydrogen (HII) regions which is basically the fraction of the IGM
volume that is occupied by ionized regions.
From its evolution ({\it bottom-middle panel}), one can see that reionization is almost completed
($Q_{\rm HII}\sim1$) around $5.8\lesssim z\lesssim7.5$ (2-$\sigma$ limits) for 
the flat Planck 2016 model.
The mean ionized fraction evolves quite rapidly, whereas the mean non-flat models favor a relatively gradual
or extended reionization starting as early as $z\approx14$. Higher the 
$\tau_{\rm el}$, the more extended is the reionization process. This is also reflected in the
evolution of the neutral hydrogen fraction $x_{\rm HI}(z)$
({\it bottom-right panel}). Here we also show various observational limits on $x_{\rm HI}(z)$ (points with different colors)
based on the measurements of quasar absorption lines, Ly$\alpha$ emitters, gamma-ray bursts (GRBs) etc.
(see Section~\ref{sec:comparison} for details). We did not include these datasets, except the most 
robust limits (open triangles) at $z\sim5-6$ from \cite{2015MNRAS.447..499M}, 
as constraints in our analyses.

We note that the TT + lowP + lensing + BAO analyses of the non-flat XCDM 
inflation model \citep{2017arXiv171003271O} and of the non-flat 
$\Lambda$CDM inflation model \citep{2017arXiv170703452O} result in almost 
identical constraints on cosmological parameter central values, with the 
central value of the XCDM equation of state parameter being $w_0 = -1$, 
This means that for this CMB and BAO dataset our non-flat $\Lambda$CDM 
reionization results also apply to the non-flat XCDM model.\footnote{Note 
that XCDM does not accurately model $\phi$CDM \citep{1988ApJ...325L..17P,1988PhRvD..37.3406R}
dark energy dynamics \citep{2000ApJ...532..109P}, so our 
reionization results here do not hold in the non-flat $\phi$CDM case.}  

\subsection{UV luminosity function}
\label{sec:lumfunc}

\begin{figure*}
\centering
  \includegraphics[height=0.28\textwidth, angle=0]{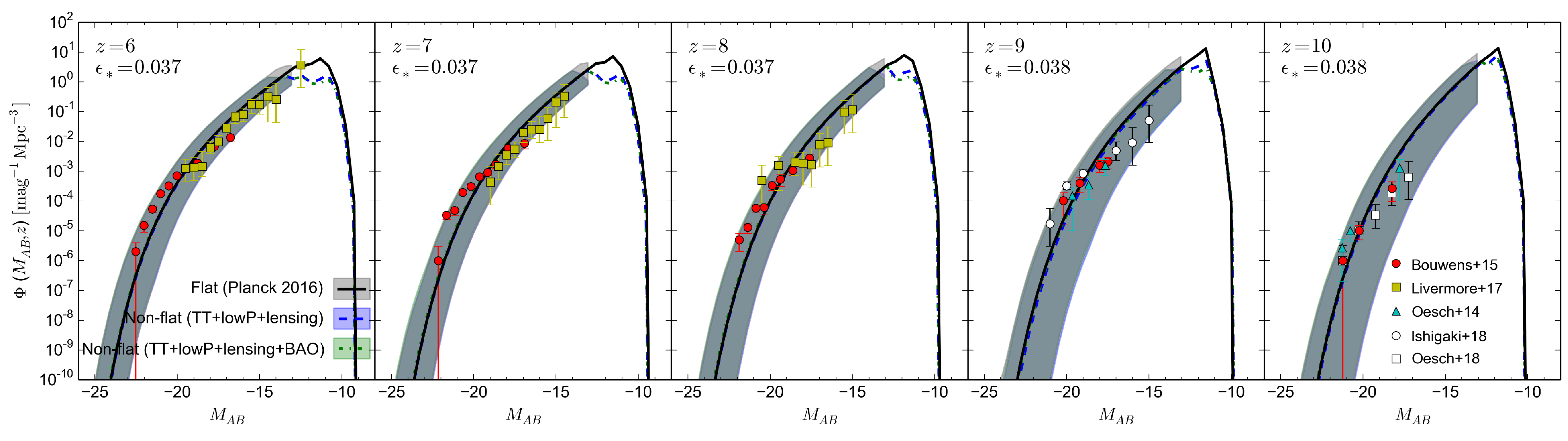}
  \caption{Evolution of high redshift ($z=6-10$) galaxy luminosity function for different models
  (flat $\Lambda$CDM and two non-flat models with and without the BAO constraints)
  with best-fit $\epsilon_*$ and the 2-$\sigma$ limits (shaded regions). The data
  points with errorbars correspond to currently available observational constraints, see the text for their
  references.}
\label{fig:lumfunc}
\end{figure*}

\begin{figure}
\centering
  \includegraphics[height=0.37\textwidth, angle=0]{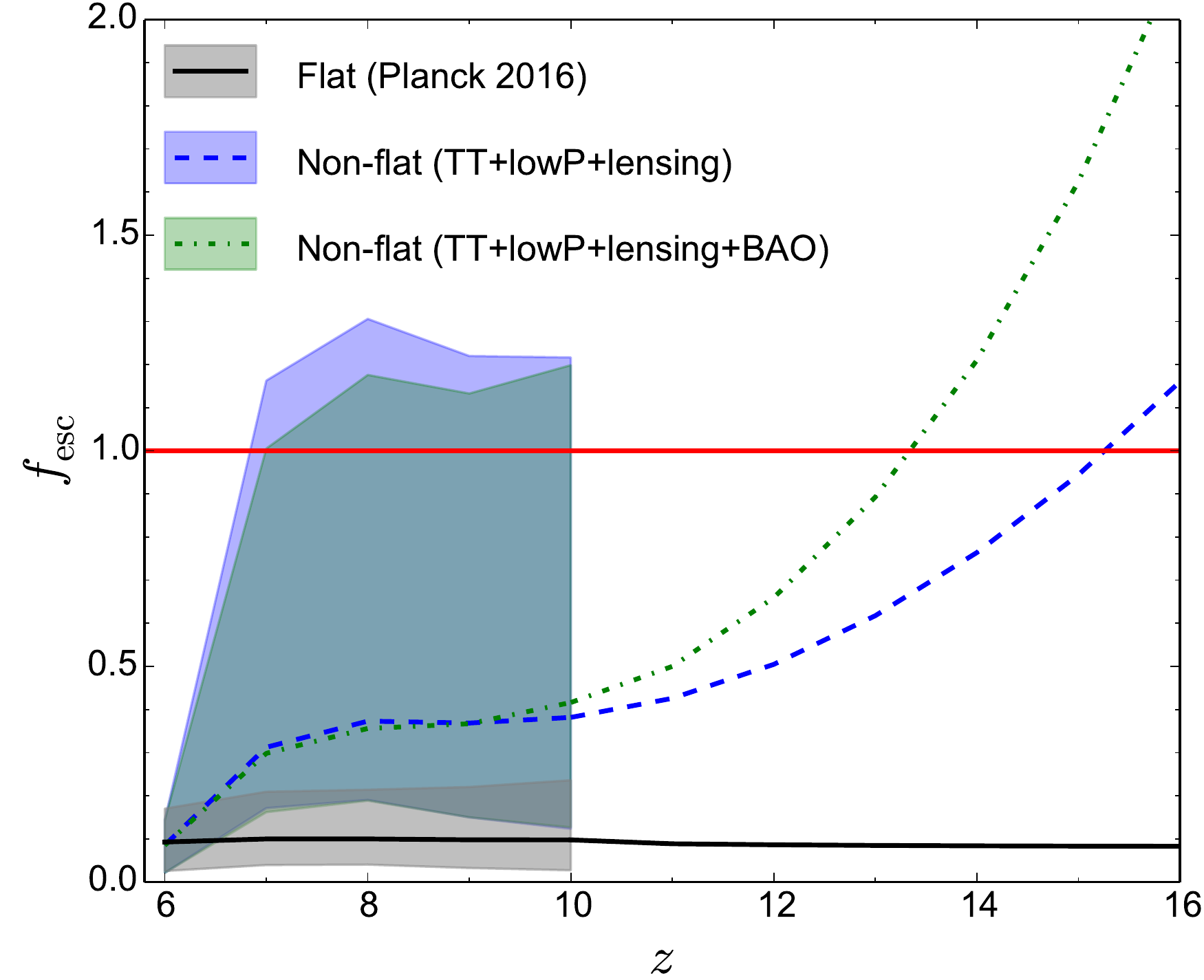}
  \caption{Redshift evolution of the escape fraction along with its 2-$\sigma$ errors for different models
  considered in this work. The red solid line indicates an escape fraction of unity.
  For $z>10$, where no observations on galaxy luminosity function exist, only the best-fit models are shown.}
\label{fig:escfrac}
\end{figure}

\begin{table*}
\centering
\begin{tabular}{@{\extracolsep{\fill} } l c c c}
Redshift & \multicolumn{3}{c}{best-fit $f_{\rm esc}$ [2-$\sigma$ C.L.]}\\
& Flat & Non-flat (TT+lowP+lensing) & Non-flat (TT+lowP+lensing+BAO)\\
\hline
\hline
$z=6$ & $0.0927$ $[0.0255,0.1702]$ & $0.0853$ $[0.0213,0.1437]$ & $0.0851$ $[0.0212,0.1434]$ \\
$z=7$ & $0.0998$ $[0.0393,0.2093]$ & $0.3125$ $[0.1713,1.1628]$ & $0.2984$ $[0.1620,1.0052]$ \\
$z=8$ & $0.0998$ $[0.0403,0.2139]$ & $0.3731$ $[0.1904,1.3059]$ & $0.3561$ $[0.1886,1.1759]$ \\
$z=9$ & $0.0979$ $[0.0324,0.2203]$ & $0.3686$ $[0.1499,1.2198]$ & $0.3669$ $[0.1500,1.1329]$ \\
$z=10$& $0.0976$ $[0.0272,0.2359]$ & $0.3817$ $[0.1233,1.2165]$ & $0.4163$ $[0.1268,1.1986]$ \\
\hline
\hline
\end{tabular}
\caption{The derived best-fit values and 2-$\sigma$ C.L. of the escape fraction for different reionization models
 at redshifts $z=6-10$.}
\label{tab:escfrac}
\end{table*}

Given the very high $N_{\rm ion}$ values and their strong evolution with redshift needed for non-flat models,
one should check whether the escape fraction needed for PopII stars becomes unrealistically high at some redshifts.
$f_{\rm esc}$ can be obtained by combining the reionization histories and the evolution of the galaxy UV
Luminosity Function (LF). This has already been studied in many of the earlier works, see e.g.,
\cite{2007MNRAS.377..285S,2009MNRAS.398.2061S,2011MNRAS.412.2781K,mitra3,mitra4}. We refer the reader to
these references for the methodology. The basic idea is to calculate the LF ($\Phi(M_{AB}, z)$; $M_{AB}$ being
the absolute AB magnitude) at redshift $z$ from the luminosity at $1500$ $\mathring{\rm A}$ of a galaxy
which depends on the star-forming efficiency of PopII stars $\epsilon_*$. We then vary $\epsilon_*$ 
as a free parameter and match the observed LFs at redshifts $z=6-10$.

In Fig. \ref{fig:lumfunc}, we present our results of the best-fit $\epsilon_*$ with $95\%$ C.L. for all three
different reionization models (indicated by the same color code as in Figure~\ref{fig:MCMC})
considered in this work. The observational datasets used here are from \cite{2015ApJ...803...34B}
for redshifts $z=6$ to $10$ (red filled circles); \cite{2017ApJ...835..113L} for galaxies at $z=6-8$
(yellow filled squares); \cite{2014ApJ...786..108O} and \cite{2018ApJ...855..105O} for redshift $9-10$ galaxy
candidates (filled cyan triangles and open squares respectively); and \cite{2018ApJ...854...73I} for $z=9$
(open circles). Although the match between data and model predictions is quite satisfactory for all redshifts
considered here, a better match can be achieved by considering a mass-dependent $\epsilon_*$ and/or
correction due to dust or halo mass quenching \citep{2010ApJ...721..193P} in the analysis which is beyond the
ambit of this paper. We find that the best-fit $\epsilon_*$ remains roughly constant ($\sim4\%$) throughout the
redshift range for all the models.

Once $\epsilon_*$ is known for different redshifts, we can obtain limits for $f_{\rm esc}$
using the MCMC constraints on the evolution of $N_{\rm ion}(z)$. Remember that, $N_{\rm ion}=\epsilon_*f_{\rm esc}N_\gamma$
where $N_{\gamma} \approx 3200$ for the PopII Salpeter IMF assumed here. The resulting $f_{\rm esc}$ values for different models
are shown in Table~\ref{tab:escfrac} and Figure~\ref{fig:escfrac}. The 2-$\sigma$ uncertainties in $f_{\rm esc}$
have been calculated using the quadrature method \citep{mitra3}.
As the star-formation efficiency is almost the same from $z=6$ to $10$, we can assume that
it will remain constant at $4\%$ even at $z>10$ and estimate the best-fit $f_{\rm esc}$ at those redshifts.
This is a reasonable assumption in the absence of galaxy luminosity function observations beyond redshift $10$.
Note that, in this figure we have shown the 2-$\sigma$ limits on $f_{\rm esc}$ only at $z=6-10$ where the
corresponding LF observables are available, whereas at $z>10$ we just extend its best-fit values
using an $\epsilon_*\approx0.04$ and best-fit $N_{\rm ion}(z)$ from the MCMC.
We find that the best-fit escape fraction remains constant at
$\sim 10\%$ for the whole redshift range in the flat $\Lambda$CDM case, whereas a strong redshift evolution of this
quantity is required for the two non-flat models -- it increases by a factor of
$\sim 5$ from $z=6$ to $10$ and approaches values as high as $\sim 100\%$ at $z\approx15$
for model without the BAO constraints (or $200\%$ in case of with-BAO model).
This also explains why the Universe is significantly ionized ($10-20\%$) even at $z\sim15$ in these non-flat
models (see the plot for $x_{\rm HI}$).
Interestingly, if we look at the 2-$\sigma$ ranges, these two non-flat models can lead to this impractically
high $f_{\rm esc}$ ($\gtrsim1$) even at redshifts $z\approx7$.
This is solely due to the fact that
$N_{\rm ion}$ for non-flat models can become as high as $\sim150$ at $z=7$, considering its 2-$\sigma$ limits
(see the {\it top-left panel} of Figure~\ref{fig:MCMC}), in order to produce such high reionization optical depths.
However, it is not possible to rule out these models based on these considerations alone as a wide range of
reionization history is still allowed at $z=7-10$ for these models due to lack of good quality
data at $z\gtrsim7$.\footnote{In addition, the smaller $\tau_{\rm el} = 0.112 \pm 0.012$
\citep{2018arXiv180305522P} found from the larger compilation of non-CMB data is about 0.7$\sigma$ smaller
than what we assume here and so will partially alleviate this tension.}       
This is also reflected in the plot of $x_{\rm HI}$ in Figure~\ref{fig:MCMC}.
In the following section, we shall see how this situation can be improved by adding constraints from LAEs
in our analysis.

A similar strong $f_{\rm esc}(z)$ evolution for higher reionization optical depths and
this striking one-to-one correspondence between them have been reported earlier.
For example, \cite{2012ApJ...746..125H} found that $f_{\rm esc}$ increases towards higher redshifts and it
becomes unity by $z\approx12$ for their {\it minimal reionization model}. They also argued that if a maximum
$f_{\rm esc}$ of $50\%$ was assumed, the same model can yield a much lower $\tau_{\rm el}=0.06$.
\cite{2012MNRAS.423..862K} claimed that a strong increase of $f_{\rm esc}$ from $\sim 4\%$ at $z=4$ to
$1$ at earlier times is needed for their reionization model to match WMAP7 $\tau_{\rm el}$ of $0.088$
(also see their Figure~5 for a direct correlation between optical depth and a constant $f_{\rm esc}$;
higher value of $f_{\rm esc}$ can result in a larger $\tau_{\rm el}$). In our earlier work \citep{mitra3}
we also found an increasing escape fraction towards higher redshifts in order to produce the desired
WMAP7 $\tau_{\rm el}$ value. However, we noted that for our model it is possible to satisfy WMAP7
and LF data simultaneously without requiring an escape fraction of order of unity at earlier epochs,
the upper limits of $f_{\rm esc}$ need be at most $50\%$ at $z=8$. 

\subsection{Inclusion of neutral fraction measurements from Ly$\alpha$ transmission at $z\sim7$}
\label{sec:LAE}

\begin{figure*}
\centering
  \includegraphics[height=0.5\textwidth, angle=0]{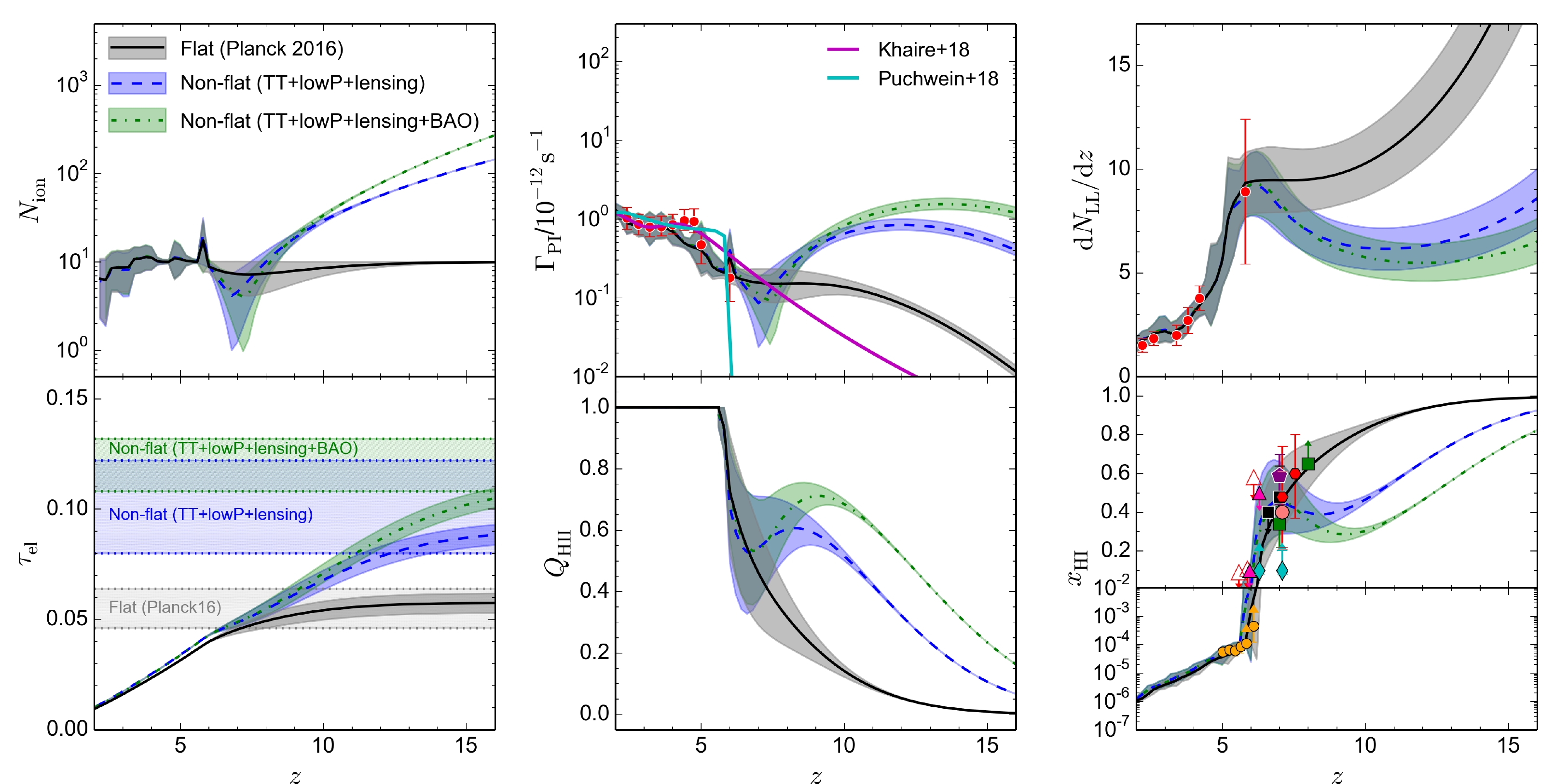}
  \caption{Same as Figure~\ref{fig:MCMC}, but now including $x_{\rm HI}$ constraint at $z\sim7$
  from \citet{2018ApJ...856....2M} indicated by filled purple pentagon in the {\it bottom-right}
  panel. See Figure~\ref{fig:xHI} for the complete references of $x_{\rm HI}$ constraints.}
\label{fig:MCMCv2}
\end{figure*}

\begin{figure}
\centering
  \includegraphics[height=0.37\textwidth, angle=0]{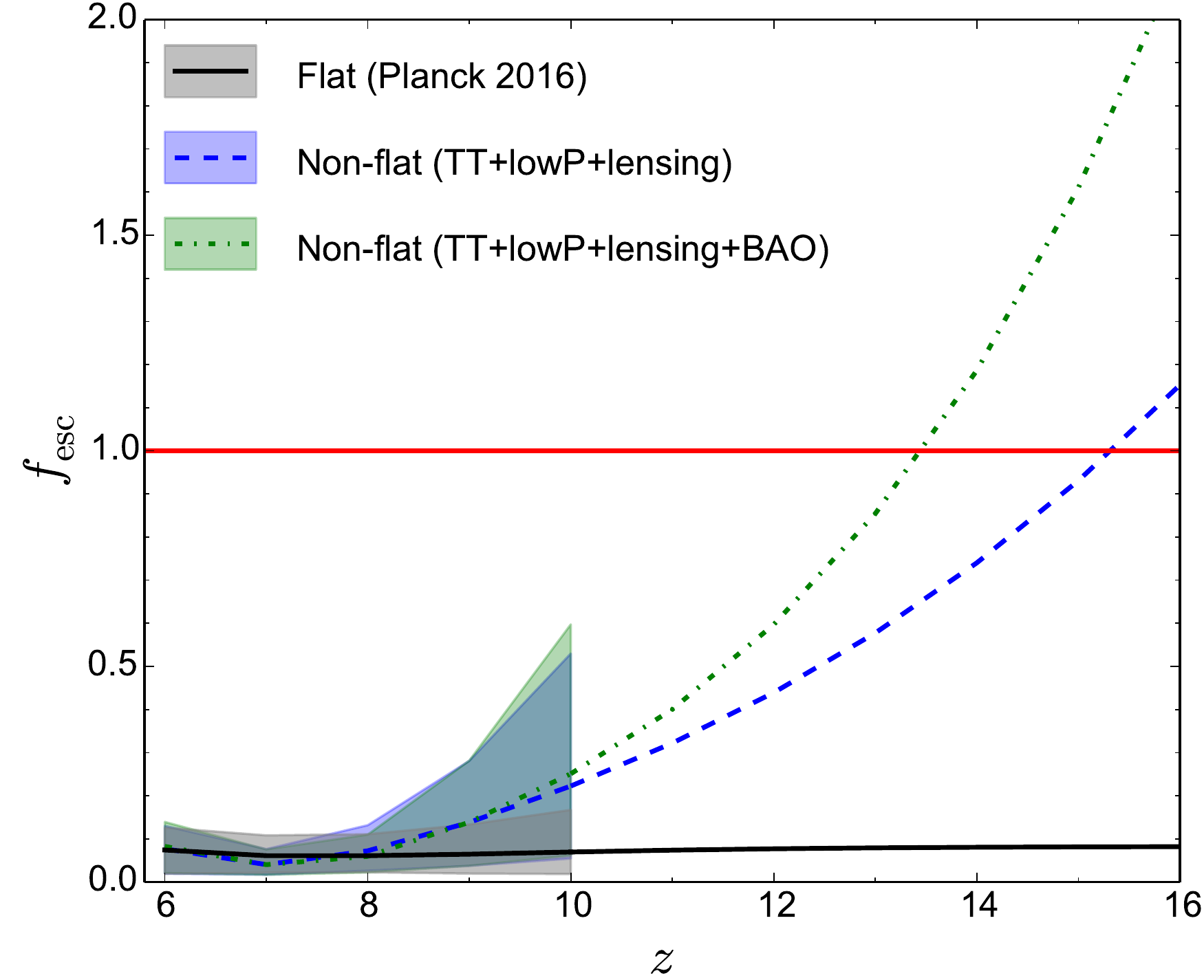}
  \caption{Same as Figure~\ref{fig:escfrac}, but now including $x_{\rm HI}$ constraint at $z\sim7$ from \citet{2018ApJ...856....2M}.}
\label{fig:escfracv2}
\end{figure}

\begin{table*}
\centering
\begin{tabular}{@{\extracolsep{\fill} } l c c c}
Redshift & \multicolumn{3}{c}{best-fit $f_{\rm esc}$ [2-$\sigma$ C.L.]}\\
& Flat & Non-flat (TT+lowP+lensing) & Non-flat (TT+lowP+lensing+BAO)\\
\hline
\hline
$z=6$ & $0.0741$ $[0.0190,0.1260]$ & $0.0771$ $[0.0193,0.1300]$ & $0.0826$ $[0.0206,0.1392]$ \\
$z=7$ & $0.0610$ $[0.0216,0.1081]$ & $0.0403$ $[0.0170,0.0762]$ & $0.0402$ $[0.0166,0.0756]$ \\
$z=8$ & $0.0610$ $[0.0225,0.1108]$ & $0.0724$ $[0.0261,0.1317]$ & $0.0595$ $[0.0223,0.1099]$ \\
$z=9$ & $0.0645$ $[0.0109,0.1333]$ & $0.1383$ $[0.0375,0.2809]$ & $0.1387$ $[0.0375,0.2816]$ \\
$z=10$& $0.0696$ $[0.0184,0.1672]$ & $0.2225$ $[0.0545,0.5299]$ & $0.2510$ $[0.0613,0.5975]$ \\
\hline
\hline
\end{tabular}
\caption{Same as Table~\ref{tab:escfrac}, but now including $x_{\rm HI}$ constraint at $z\sim7$ from \citet{2018ApJ...856....2M}.}
\label{tab:escfracv2}
\end{table*}

So far the reionization histories at $z>6$ depend only on the value of $\tau_{\rm el}$ coming from
CMB observations, and thus the constraints remain relatively weaker at those redshifts. One can,
in principle, include other high-redshift non-CMB datasets in order to further strengthen the model
constraints.
We have indicated some of those possibilities in the plot for $x_{\rm HI}$
(or see Section~\ref{sec:comparison} for details).
Although these data are highly model-dependent and might get modified in the future, it would be interesting
to check if the constraints improve significantly by including such measurements available at $z>6$.
To this end, here we have included one more observable, the constraint on the global neutral fraction
at $z\sim7$ of $x_{\rm HI}=0.59^{+0.11}_{-0.15}$ from \cite{2018ApJ...856....2M}, in addition
to the earlier datasets mentioned in Section~\ref{sec:cfmodel}. This data is inferred from
a sample of observed Lyman Break galaxies (LBGs) presented in \cite{2014ApJ...793..113P} using
a Bayesian inference framework and sophisticated IGM simulations.

The resulting reionization constraints are shown in Figure~\ref{fig:MCMCv2}. The first thing to note is
that the 2-$\sigma$ limits are considerably reduced for all the models considered here since we now
force the model to match the $x_{\rm HI}$ constraint. Such high value of $x_{\rm HI}$ at $z=7$
essentially disfavours a large set of models which were otherwise allowed in our earlier analysis.
Although we still require a similar strong redshift evolution of $N_{\rm ion}$ for the non-flat cases,
the rise is rather late, starting at $z>8$ and then rapidly increasing towards higher redshifts.
This indicates that the PopII stars dominate the reionization over a longer period of time
up to $z\approx8$ and after that a sharp increase in photon escape fraction and/or the
PopIII stars take over, so that enough contribution to $\tau_{\rm el}$ is acquired
to match its corresponding value. However, all these models produce somewhat lower $\tau_{\rm el}$
than what we got earlier, reflecting a possible tension between high-$z$ LAE data and a very
large optical depth value ($\gtrsim 0.11$). In fact, we find that the best-fit $\chi^2$-values
increase by $\sim 2.5$ for the non-flat models when the additional LAE constraint is included
in the analysis (the corresponding rise in the best-fit $\chi^2$ is only $0.7$ for the flat model).
This indicates that the non-flat models tend to perform worse in presence of the LAE constraint,
however, they cannot still be conclusively ruled out because of the large error-bars in the data.

The constraints on $Q_{\rm HII}$ and $x_{\rm HI}$
change significantly in this case, in particular they become very constricted near the end-stage
of reionization. Reionization is almost completed ($Q_{\rm HII}\sim1$) at $5.8\lesssim z\lesssim6.0$
(2-$\sigma$ limits) irrespective of the model we choose. Hence, if we include the $x_{\rm HI}$
measurements at $z\gtrsim7$ in the analysis, completion of reionization cannot occur earlier than
$z\approx6$, essentially ruling out most of the models of early reionization which were allowed
previously. The growth of $Q_{\rm HII}$ is gradual for the flat model. On the other hand, the non-flat
models, which are characterized by a sharp rise in $N_{\rm ion}$ and $\Gamma_{\rm PI}$ at $z>8$
in order to produce high optical depths, indicate a much faster increase in $Q_{\rm HII}$ at
initial stages, followed by a sharp fall around $z\approx8$ (corresponding to a sharp decrease
in $N_{\rm ion}$) to match the $x_{\rm HI}$ measurements (filled purple pentagon in the plot). 
Similar conclusions can be obtained from the plot of neutral fraction. For the non-flat models,
it shows a gradual decrease up to $z\sim8$ from its higher value at earlier redshifts, then a rapid increase 
up to $z\approx7$ in order to obey the observed $x_{\rm HI}$ limit and finally it decreases
again to smoothly match the Ly$\alpha$ forest data at $z\lesssim6$.

Although the inclusion of LAE data reduces the tension between the predicted $x_{\rm HI}$
from non-flat models with the observed value at $z\sim7$, it now introduces a clear non-monotonic
redshift evolution of photoionization rates, LLSs and the neutral fraction. It will be very
difficult for any physically motivated reionization model\footnote{Perhaps a model having
an abrupt transition from PopIII to PopII stars around redshifts 7-8 (e.g. a step function of $N_{\rm ion}$;
\citealt{mitra1}) might explain such non-monotonic evolution.} to justify such trends like the sudden
decrease in LLSs at $z>6$ (corresponding to an abrupt increase in mean-free path of ionizing photons
around this redshift)
or the recombination of hydrogen again at $z\sim6-9$. Also such rapid boost in the evolution of
photoionization rates at $z>7$ lacks a meaningful explanation and cannot be naturally produced
by any UV background model. For comparison we have plotted here the $\Gamma_{\rm PI}$ evolution
predicted by two recent models from \cite{2018arXiv180109693K} (solid purple curve) and
\cite{2018arXiv180104931P} (cyan curve) which clearly shows the opposite trends. In fact,
the monotonic evolution of $\Gamma_{\rm PI}$ from the flat CDM case is somewhat more agreeable
with their results.

Next we try to fit the observed UV luminosity functions at $z=6-10$ following the same method described in
Section~\ref{sec:lumfunc}, and the results are almost similar to those obtained from our previous
analysis (Figure~\ref{fig:lumfunc}). The same non-evolving SF efficiency of $\epsilon_*\approx4\%$
is required for all redshift ranges. However the constraints on $f_{\rm esc}$ at $z=6-10$ modify considerably
(shown in Figure~\ref{fig:escfracv2} and Table~\ref{tab:escfracv2}) due to the change in
$N_{\rm ion}$. The escape fraction remains unchanged at $\approx10\%$ for $z\lesssim8$ in all the
models reflecting the non-evolving nature of $N_{\rm ion}$ at those redshifts. Then it increases
moderately towards redshift $z\approx10$ for non-flat cases. In particular, for the model with highest
$\tau_{\rm el}$ a maximum of $60\%$ (2-$\sigma$ C.L.) photon escape fraction is needed at
redshift $10$. Unlike the previous case, now the non-flat models do not require unrealistically
high value of $f_{\rm esc}$ at $z\lesssim10$. This is because the inclusion of the $z\sim7$ $x_{\rm HI}$
constraint forces these models to a relatively late
reionization scenario reducing the need of significantly higher amount of early epoch
($6\lesssim z\lesssim10$) sources or very large PopII escape fractions. But we still need
the best-fit $f_{\rm esc}$ to be $\gtrsim1$ at higher redshifts, as a very high $N_{\rm ion}$
($>100$) at $z\gtrsim13$ is required for these non-flat models to match the corresponding optical
depth constraints.

Nonetheless, we should mention that the inferred $x_{\rm HI}$ measurements from high-$z$ LAEs
can have model dependencies and large uncertainties, e.g., effects of dust extinction
\citep{dayal1}, self-shielded absorbers \citep{2013MNRAS.429.1695B,2015MNRAS.452..261C,
2016MNRAS.463.4019K,2018arXiv180303789W}, or infall of circumgalactic medium (CGM) matter in the haloes
\citep{2017ApJ...839...44S,2018arXiv180303789W}. Thus any conclusions drawn by incorporating
them in our analysis should be interpreted with caution.

\subsection{Comparison of the results with other data}
\label{sec:comparison}

\begin{figure}
\centering
  \includegraphics[height=0.37\textwidth, angle=0]{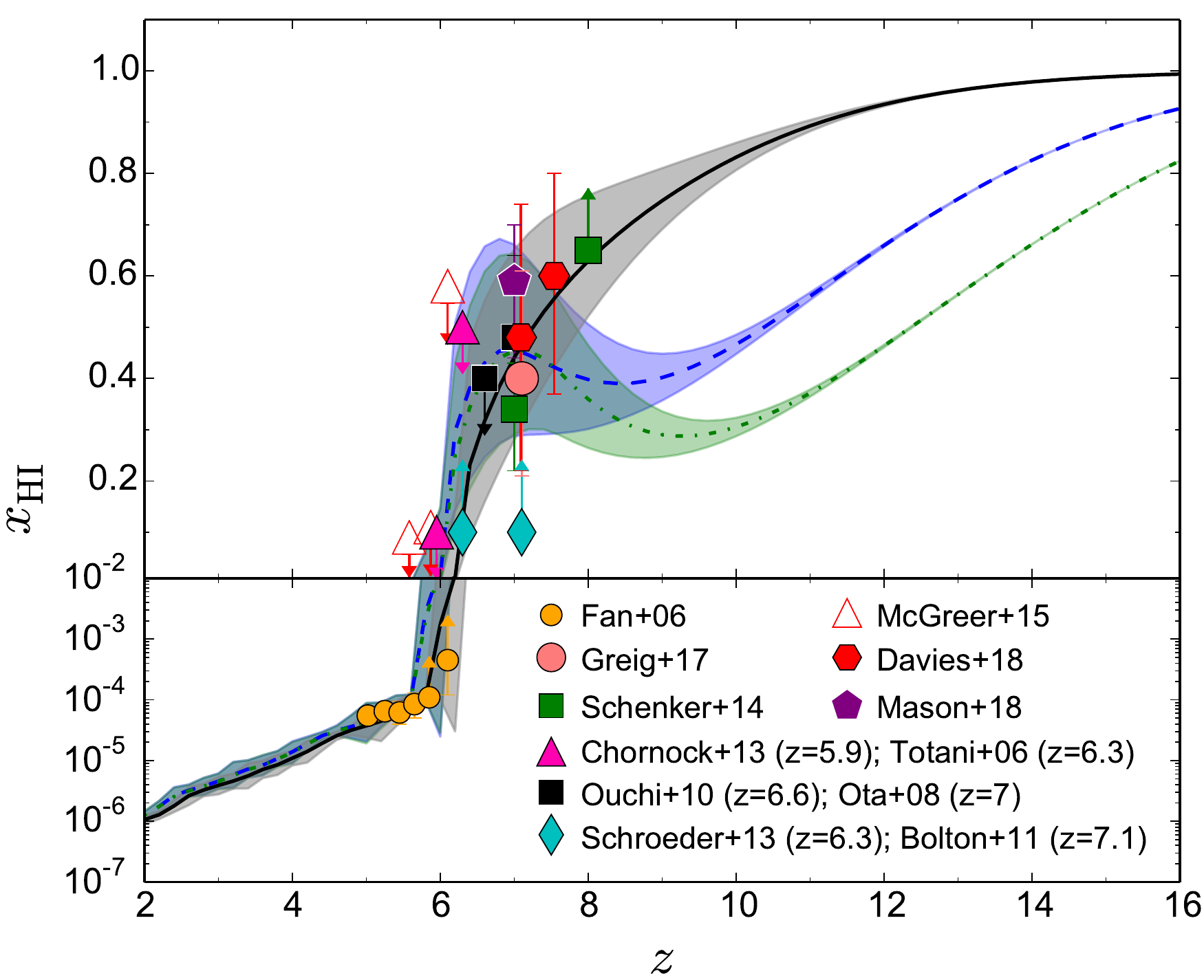}
  \caption{{\it Bottom-right} panel of Figure \ref{fig:MCMCv2} -- evolution of 
  the neutral hydrogen fraction compared with various existing observations listed here and described
  further in the text.}
\label{fig:xHI}
\end{figure}

Finally, we focus on the comparison of our model predictions for the neutral fraction
($x_{\rm HI}$) with other data. A separate plot for $x_{\rm HI}$ from the analysis presented
in the last section is shown in Figure~\ref{fig:xHI} (same as the {\it bottom-right}
panel of Figure~\ref{fig:MCMCv2}) for clarity. Although the majority of the data shown 
in this figure provide weak and model-dependent constraints on 
the EoR, it is instructive to compare our model predictions with these data.

\begin{itemize}
 \item {\it Quasars:} The strongest evidence related to reionization perhaps comes from the Ly$\alpha$ forest
 data of high-redshift quasars from \cite{2006AJ....132..117F}. However, estimating
 the volume-averaged neutral fraction from the original data involves adoption of a particular model of
 IGM density distribution and temperature evolution. From the evolution of $x_{\rm HI}(z)$,
 one can immediately see a striking match between all our models and these data,
 shown here by filled yellow circles, even though we did not use these data in 
 our MCMC analyses.
 This is not unexpected for the following reason. \cite{2006AJ....132..117F} 
 assume a simple parametric form for the density distribution function \citep{2000ApJ...530....1M}
 which is qualitatively very similar to the lognormal 
 distribution adopted here \citep{mitra4}. Also their IGM inhomogeneities
 are calculated from the evolution of the mean free path using the 
 same \citet{2000ApJ...530....1M} prescription we use in our model.
 Other evidence comes from the observations of quasar near zones.
 Bright quasars at early epochs ($z\sim6-7$) can create the largest ionized regions around them, known as
 near zones, and thus can have a prominent effect on the IGM at the tail-end 
 of reionization \citep{2007MNRAS.381L..35B,2010ApJ...714..834C,2014MNRAS.443.3761P}. Recent measurements of these
 at $z\sim6.3$ by \cite{2013MNRAS.428.3058S} and at $z\sim7.1$ by \cite{2011MNRAS.416L..70B} infer
 a corresponding lower limit on mean $x_{\rm HI}\gtrsim0.1$ (filled cyan diamonds in the figure).
 More recently, \cite{2017MNRAS.466.4239G} and \cite{2018arXiv180206066D} constrained the neutral
 fraction from the damping wing analysis of highest redshift ($z>7$) quasars known. The mean $x_{\rm HI}=0.40^{+0.21 }_{-0.19}$
 at $z=7.09$ from \cite{2017MNRAS.466.4239G} and $x_{\rm HI}=0.60^{+0.20}_{-0.23} (0.48^{+0.26}_{-0.26})$
 at $z=7.54 (7.09)$ from \cite{2018arXiv180206066D} are shown here by filled salmon circle
 and red hexagons respectively. However, there
 could be several ambiguities in estimating these constraints due to our poor understanding of the intrinsic
 properties of observed quasars \citep{2007MNRAS.374..493B,2007MNRAS.376L..34M}, and hence these have
 not been used here for constraining our model parameters.
 More useful constraints for us instead
 come from a model independent dark pixel analysis of high-$z$ quasar spectra by \cite{2015MNRAS.447..499M},
 especially the upper limits at $z\sim5.6$ and $5.9$ (open red triangles). We ensure our models not
 defy these bounds by imposing a prior in the MCMC analysis, which guarantees 
 that reionization is almost completed at least by redshift $\sim5.8$.
 
 \item {\it Gamma-ray bursts:} The afterglow spectra of gamma-ray bursts (GRBs) is another potential probe
 of the EoR \citep{2006ApJ...642..382B}. We show the constraints from observed GRB host galaxies of $x_{\rm HI}\lesssim0.5$
 at $z\sim6.3$ \citep{2006PASJ...58..485T} and $x_{\rm HI}\lesssim0.1$ at $z\sim5.9$ \citep{2013ApJ...774...26C}
 by filled pink triangles. Although these data are relatively weak due to the intrinsic damped Ly$\alpha$
 absorption, predictions from all our models, interestingly, quite reasonably 
 obey these limits.
 
 \item {\it Ly$\alpha$ emitters:} As the number densities of observed quasars and GRBs decline at high redshifts,
 one must look at the next higher-redshift reliable probe of the EoR, the Ly$\alpha$ emitters (LAEs).
 Studies of Lyman$\alpha$ emitting galaxies near the end of the EoR have proven crucial for understanding
 reionization processes, because of the attenuation of Ly$\alpha$ emission lines by neutral contents left in the IGM at this epoch
 \citep{2009ApJ...696.1164O}. Observations of LAEs at $z=6.6$ by \cite{2010ApJ...723..869O} and at $z=7$ by
 \cite{2008ApJ...677...12O} infer the values of $x_{\rm HI}$ to be $\lesssim0.4$ and $0.32-0.64$ respectively
 (shown in the plot by filled black squares). More recently, \cite{2014ApJ...795...20S} have presented the most promising
 measurements of Ly$\alpha$ emission at the highest redshift known and provide an estimate of neutral
 fraction to be $x_{\rm HI}=0.34^{+0.09}_{-0.12}$ at $z\sim7$ and $x_{\rm HI}>0.65$ at $z\sim8$ (filled green squares).
 The resulting 2-$\sigma$ MCMC limits on this quantity from our non-flat models seem to be significantly low at this
 redshift; it can take values at most $\sim0.5$ at $z=8$.
\citet{2015MNRAS.452..261C}, using simulations of the high-redshift IGM, showed that the evolution
in the LAE number density at $z \gtrsim 6.6$ is in better agreement with reionization models having $\tau_{\rm el} \lesssim 0.066$.
 Even though there might exist several uncertainties in the estimation of $x_{\rm HI}$ and reionization history from LAE data, we can
 say that the most severe challenges for the non-flat models come from
these datasets.
 On the other hand, the lower $\tau_{\rm el}$ data for flat model makes it possible to produce a moderate
 evolution of $x_{\rm HI}$ in agreement with the current observed limits for all redshifts.
\end{itemize}

 The fact that a model with a higher reionization optical depth produces
 a considerably smaller neutral fraction at earlier times has been reported earlier
 \citep{2013ApJ...768...71R,2015ApJ...802L..19R,2015ApJ...811..140B,mitra4}. In particular, \cite{2015ApJ...802L..19R}
 demonstrated that it is possible to simultaneously match the lower $\tau_{\rm el}$ from Planck 2015
 and most of the observed constraints on $x_{\rm HI}$ in the range $6\lesssim z\lesssim8$ using the
 latest Hubble Space Telescope data on the star formation rate density $\rho_{\rm SFR}(z)$.
 However a model with a higher $\tau_{\rm el}$ (e.g. $0.088$ from nine years of
 Wilkinson  Microwave Anisotropy Probe or WMAP9 observations) would require a 
dramatic increase of SFR at $z\gtrsim7.5$ and hence lead to a notable 
inconsistency with several observations on the neutral fraction. In fact, a 
very similar trend can also be seen in their earlier work
 \citep{2013ApJ...768...71R} where they showed that a model that matches the 
observed $x_{\rm HI}$ quite well struggles to produce such a large WMAP9 
$\tau_{\rm el}$ value.

\section{Concluding remarks}
\label{sec:conclusions}

We have presented a detailed statistical analysis of reionization in closed
$\Lambda$CDM inflation models using joint datasets of CMB and quasars. In 
particular, we compare how reionization proceeded over cosmic time in the 
flat and non-flat models. In the non-flat models under consideration the 
cosmological parameters are constrained by the Planck 2015 CMB data (also in 
combination with the BAO measurements) using a consistent energy density 
inhomogeneity power spectrum \citep{2017arXiv170703452O}. These data prefer 
mildly closed ($\Omega_{\rm k}<0$) models with the curvature density 
parameter contributing only $1\%-2\%$ of the total mass-energy budget of 
the Universe. Such models not only reasonably match many observations but 
might also improve the agreement with observed low-$\ell$ $C_\ell$'s and
weak lensing determined $\sigma_8$'s, although they do somewhat worsen the
high-$\ell$ $C_\ell$ fit. However, these models predict a 
relatively higher reionization optical depth than that found from Planck 
2016 data with the spatially-flat tilted $\Lambda$CDM model. This could 
result in a completely different reionization history at earlier epochs ($z>6$)
in the non-flat cases.

Our main results, in summary, are:
\begin{itemize}
 \item We find all three models behave the same way in the lower redshift 
regime ($z\lesssim6$), as expected,  whereas their predictions at higher-$z$
depart from each other due to the differences in optical depth values.
 \item Unlike the flat case, the non-flat models need many more high redshift
 reionization sources. A changing IMF influenced by PopIII stars and a 
strong evolution in  the photon escape fraction of galaxies are two 
possibilities.
 \item For the usual flat model from Planck 2016, the lower optical depth 
favors a relatively quicker evolution of reionization. On the other hand, 
a more gradual or extended reionization is found for the non-flat models. 
In fact, larger the optical depth, more gradual is the reionization process.
 \item The resulting neutral hydrogen fraction seems to be quite small at 
higher redshifts ($z>7$) for the non-flat models compared to the flat one. 
Such small values, e.g. $\lesssim0.4$ at $z\sim8$, are likely disfavored 
by current observational bounds from distant Ly$\alpha$ emitters.
 \item This also reflects in the evolution of escape fraction. $f_{\rm esc}$
 must be higher at earlier epochs for the non-flat models. 
 The best-fit  $f_{\rm esc}$ increases from $\sim10\%$ at $z=6$ to $\sim40\%$ at $z=10$,
 and up to $>100\%$ at much higher redshifts,
 and considering its 2-$\sigma$ limits it can become unrealistically high
 ($>1$) even at $z\gtrsim7$. On the other hand, a constant escape fraction
 of $\sim10\%$ is sufficient for the flat $\Lambda$CDM model.
\end{itemize}

One can see that, apart from the constraint on $x_{\rm HI}$ at $z\sim8$ (filled 
square lower limit point in {\it bottom-right panel} of Fig.~\ref{fig:MCMC}), 
the non-flat models, considering their 2-$\sigma$ limits, are not yet in 
conflict with most of the measurements related to reionization. The $z\sim8$
data comes from the recent estimates of evolving LAEs by \cite{2014ApJ...795...20S}. Those observational
results are then converted to $x_{\rm HI}$ by adopting a suitable model appropriate for patchy reionization
\citep{2007MNRAS.381...75M,2012ApJ...744..179S}. This conversion however involves modeling several
uncertain key parameters, like the escape fraction of 
ionizing photons, the degree of self-shielding etc. and by necessity this 
will bring in model dependencies. In fact, most of the observed $x_{\rm HI}$ 
constraints at $z\gtrsim7$ are somewhat model dependent, hence we 
did not include them in the main MCMC analysis.
Nevertheless, in order to examine how the non-flat models perform if one uses such
data to constrain the reionization history, we later included the observed
$x_{\rm HI}$ at $z\sim7$ from \cite{2018ApJ...856....2M} keeping in mind that
the results might be significantly biased by uncertainties in interpreting the data.
The resulting 2-$\sigma$ limits at $z>6$ now reduce considerably for all the
models due to this additional high redshift data. For non-flat scenarios
$N_{\rm ion}$ remains constant up to $z\approx8$, then increases rapidly
at higher redshifts. In fact it behaves somewhat similar to the lower
bound of $N_{\rm ion}$ plotted in Figure~\ref{fig:MCMC}, which signifies
that the PopII stars remain dominating until $z\sim8$ in order to match
the $x_{\rm HI}$ constraint included here. As a result we get an almost
constant $f_{\rm esc}$ of $\sim10\%$ up to $z=8$ with moderately increasing
(maximum of $60\%$ for 2-$\sigma$ limits) towards $z=10$, indicating that the non-flat
models are still permitted by the LBG data at $z\sim7$.
However if we continue to higher redshifts, assuming the same constant
$4\%$ SF efficiency, where no actual observations on galaxy LF exist,
the best-fit $f_{\rm esc}$ can again become unrealistically high
for these models. Also, the evolution of various reionization quantities
(e.g. photoionization rate, LLSs, neutral fraction etc. from Figure~\ref{fig:MCMCv2}) becomes significantly
non-monotonic in nature, especially when we include the LAE data. It will not be straightforward for
any physical model to account for such trends.
Interestingly, we find that the non-flat models perform much worse in terms of the
best-fit $\chi^2$ when the LAE constraints are included, however, the error-bars are
not small enough to rule them out.

Although it is now well understood that the LAE data prefer a late reionization
\citep{2015MNRAS.446..566M,2015MNRAS.452..261C} and the non-flat models 
struggle to match this, we still probably have to rely on upcoming 
observations on high-redshift reionization sources to conclusively
rule out the non-flat models. Finally, we end this paper by indicating 
some, likely to be decisive, future observational prospects in this regard.

In the next few years there will be excellent openings on various 
observational fronts for greatly improving our understanding of the end phases 
of the EoR. Future observations of more high-redshift quasars are expected
to come from the Large Synoptic Survey Telescope (LSST)\footnote{https://www.lsst.org/},
Euclid\footnote{https://www.euclid-ec.org/}, the Wide-Field Infrared 
Survey Telescope (WFIRST)\footnote{https://wfirst.gsfc.nasa.gov/}, the Thirty 
Meter Telescope (TMT)\footnote{http://www.tmt.org/} and the Giant Magellan 
Telescope (GMT)\footnote{https://www.gmto.org/}, which can significantly 
increase our knowledge on the timing and nature of reionization. Furthermore, 
the James Webb Space Telescope (JWST)\footnote{https://www.jwst.nasa.gov/}, 
the Atacama Large Millimeter Array (ALMA)\footnote{http://www.almaobservatory.org}
and the Hyper Suprime-Cam (HSC)\footnote{https://www.naoj.org/Projects/HSC/} 
on the Subaru telescope seem to be most promising instruments to target 
high-redshift LAEs as a very powerful reionization probe. And finally,
the detection of the redshifted 21-cm signal from the EoR by several radio 
telescopes like the Giant Metrewave Radio Telescope
(GMRT)\footnote{http://www.gmrt.ncra.tifr.res.in/}, the Murchison Widefield Array
(MWA)\footnote{http://www.mwatelescope.org/}, the Hydrogen Epoch of Reionization Array
(HERA)\footnote{http://www.reionization.org/} and the Low-Frequency Aperture Array
(LFAA) of the Square Kilometre Array (SKA)\footnote{https://www.skatelescope.org/}
will provide direct probes of the ${\rm HI}$ distribution in the diffuse 
IGM, which should be able to adjudicate between the different
reionization scenarios of the flat and non-flat models.

\section*{Acknowledgements}

We thank G.\ Holder for valuable comments. B.R.\ is supported
in part by DOE grant DE-SC0011840.

\bibliography{reionization-smitra}
\bibliographystyle{mnras}

\end{document}